\documentclass{iopart}
\usepackage{amssymb}
\usepackage{graphicx}
\usepackage{psfrag}

\begin{document}

\letter{Decoherence bypass of macroscopic superpositions in quantum  measurement} 

\author{Dominique Spehner$^1$ and Fritz Haake$^{2}$}

\address{$^1$Institut Fourier, 100 rue des Maths, 38402
  Saint-Martin d'H\`eres, France}

\address{$^2$Fachbereich Physik,
  Universit{\"a}t Duisburg-Essen, 47048 Duisburg, Germany }

\date{\today}

\eads{spehner@ujf-grenoble.fr}

\begin{abstract}
  We study a class of quantum measurement models. A microscopic object
  is entangled with a macroscopic pointer such that a distinct pointer
  position is tied to each eigenvalue of the measured object
  observable.  Those different pointer positions mutually decohere
  under the influence of an environment.  Overcoming limitations of
  previous approaches we (i) cope with initial  correlations
  between  pointer and environment by considering them
  initially in a metastable 
  local thermal equilibrium, (ii) allow for object-pointer
  entanglement and environment-induced decoherence of distinct pointer
  readouts to proceed simultaneously, such that mixtures of
  macroscopically distinct object-pointer product states arise without
  intervening macroscopic superpositions, and (iii) go beyond the
  Markovian  treatment of decoherence.
\end{abstract}






\newcommand{\I}{{\rm{i}}}
\newcommand{\D}{{\rm{d}}}
\newcommand{\E}{{\rm{e}}}


\newcommand{\Sc}{{\cal S}}
\newcommand{\Pc}{{\cal P}}
\newcommand{\Bc}{{\cal B}}
\newcommand{\ket}[1]{| #1 \rangle}
\newcommand{\bra}[1]{\langle #1 |}
\newcommand{\braket}[2]{\langle #1 | #2 \rangle}
\newcommand{\ketbra}[2]{| #1 \rangle \langle #2 |}
\newcommand{\meanB}[1]{\langle #1 \rangle }
\newcommand{\meanBx}[2]{\langle #1 \rangle_{#2 } }
\newcommand{\lamb}{\lambda}
\newcommand{\HS}{{H}_{\cal{S}} }
\newcommand{\HB}{{H}_{\cal{B}} }
\newcommand{\HP}{{H}_{\cal{P}} }
\newcommand{\HPS}{{H}_{\cal{PS}} }
\newcommand{\HtildePS}{\widetilde{H}_{\cal{PS}} }
\newcommand{\HPB}{{H}_{\cal{PB}} }
\newcommand{\HPSB}{{H} }
\newcommand{\rhoS}{ {\rho}_{\cal{S}} }
\newcommand{\rhoSfree}{ {\rho}_{\cal{S}}^{(0)} }
\newcommand{\rhoPfree}{ {\rho}_{\cal{P}}^{(0)} }
\newcommand{\rhoP}{ {\rho}_{\cal{P}} }
\newcommand{\rhoPeq}{ {\rho}_{\,\cal{P}}^{(\rm{eq})} }
\newcommand{\rhoB}{ {\rho}_{\,\cal{B}}^{(\rm{eq})} }
\newcommand{\rhoPS}{ {\rho}_{\cal{PS}} }
\newcommand{\rhoPB}{ {\rho}_{\,\cal{PB}}}
\newcommand{\rhoPBeq}{ {\rho}_{\,\cal{PB}}^{(\rm{eq})} }
\newcommand{\rhoPSB}{ {\rho} }
\newcommand{\trB}{\tr_{\mathcal B}}
\newcommand{\TS}{T_{\mathcal{S}} }
\newcommand{\TP}{T_{\mathcal{P}} }
\newcommand{\TB}{T_{\mathcal{B}} }
\newcommand{\TBmin}{t_{\mathcal{B}} }
\newcommand{\ZB}{Z_{\mathcal{B}} }
\newcommand{\ZBx}[1]{Z_{#1} }
\newcommand{\ZPB}{Z_{\cal{P B}} }
\newcommand{\ds}{\delta s }
\newcommand{\dectime}{t_{\rm dec} }
\newcommand{\enttime}{t_{\rm ent} }
\newcommand{\club}{${\clubsuit}$}


\section{Introduction}

The interpretation and theoretical description of
measurements on quantum systems have been under debate since the birth
of quantum theory~\cite{Wheeler}.  More recently, interest for
this question has been revived by new developments in quantum
information. Quantum detection can be used either to extract
information on quantum states or  to monitor quantum systems (quantum
trajectories~\cite{Knight98}, quantum Zeno effect~\cite{Fischer01}).
Quantitative
treatments of measurement models serve both to elucidate the
self-consistency of quantum theory and its interpretation, and to
calculate the time scales relevant for experiments.  Data for the
decoherence time are in fact accumulating, in microwave
cavities~\cite{Brune96}, in solid state devices like superconducting
tunnel junction nanocircuits~\cite{Vion02,Buisson06}, and in electron beams
interacting with a semiconducting plate~\cite{Sonnentag}.
On a more fundamental ground, ever larger
classes of nonlocal hidden-variable theories are being ruled out
experimentally as competitors
of quantum mechanics~\cite{Zeilinger}. 

We propose in this letter to extend the range of validity of the approach
by Zeh, Zurek, and others based on environment-induced decoherence
\cite{Giulini,Zurek91,Zurek03,Haake93,Balian01,Balian03}.  In this vein, i.e., on the
basis of the probabilistic interpretation of quantum mechanics (Born
rules) and of a lack of knowledge about the microscopic degrees of
freedom of the apparatus, we demonstrate that the reading
of a macroscopic pointer of the apparatus reveals an eigenvalue of
an observable of the measured quantum object, in spite of the unitary
evolution of the composite system (object and apparatus).  We obtain
explicit results for the object-pointer dynamics for a class of
models under realistic assumptions.
We would like to note that macroscopic manifestations of microscopic quantum
fluctuations are not the exclusive privilege of measurement:
they also appear e.g. in  superfluorescence, where intense light
pulses display substantial shot-to-shot fluctuations~\cite{supfluor}
and in  cosmic rays, where single particles propagating in 
a fluctuating spacetime generate large
showers of particles in the atmosphere.

During an ideal  measurement the quantum object ($\Sc$)
interacts with the pointer ($\Pc$) in  such a way that a one-to-one
correspondence arises between the eigenvalues 
of the measured object observable  
and macroscopically distinct pointer states.  The coupling of $\Pc$
with its environment (``bath'' $\Bc$) causes decoherence.
Most previous work  deals with 
these two interactions separately  
(see, however, \cite{Balian01,Balian03}):
a first step
(``premeasurement'') exclusively treats the unitary evolution
entangling $\Sc$ and $\Pc$ and  yields a superposition of the
macroscopically distinct object-pointer states associated with each
eigenvalue of the measured observable. That latter
``Schr\"odinger cat'' state is taken as the initial state for a second
process, decoherence; there, the quantum
correlations between object and apparatus are transformed into
classical correlations, as the superposition of object-pointer states
is degraded to a statistical mixture of the same states. For such a
sequential treatment to make physical sense, the time duration 
of the entanglement process would have to be short compared
with the decoherence time $t_{\rm dec}$. But since the latter tends to
be very small for macroscopic superpositions, that assumption is quite
questionable. A second shortcoming of some previous work lies in
the assumption of initial statistical independence of
$\Pc$ and $\Bc$; since these two systems cannot be isolated from 
each other,
a more realistic assumption is
a metastable local thermal state of $\Pc+\Bc$. 
As a third restriction, memory effects are often
neglected for the 
(reduced) object-pointer dynamics. 
 That Markov approximation assumes
$\dectime$ to be larger than the bath correlation time $\TB$, a
condition not satisfied
in some experiments~\cite{Vion02,Buisson06}
and of questionable validity  
 if macroscopic
or even mesoscopic superpositions or mixtures
arise during the object-pointer evolution.  

We overcome the three aforementioned deficiencies.
The key is an assumption
for a certain ordering of time scales: the decoherence time 
$t_{\rm  dec}$ and the object-pointer interaction time $t_{\rm int}$
must be small compared to the characteristic time $\TS$ 
($\TP$)
of the evolution of the  measured observable $S$ 
(the pointer position $X$) under the object Hamiltonian 
$\HS$ (the pointer
Hamiltonian $\HP$), 
\begin{equation}\label{timescales}
t_{\rm dec}, t_{\rm int} \ll \TS 
\qquad , \qquad t_{\rm dec}, t_{\rm int} , \hbar\beta \ll \TP\,.
\end{equation}
In the last limit, $\beta=(k_B T)^{-1}$ denotes the inverse temperature of the bath.
As desirable for a measurement, the free dynamics of
$\Sc$ then remains ineffective on $S$  
during the measurement, i.e., 
$S^0(t)= \E^{\I  \HS t/\hbar} S \E^{-\I 
  \HS t /\hbar}\simeq S$ for
$t \lesssim \dectime, t_{\rm int}$. Here  
$S^0(t)\simeq S$ means that
the expectation value of 
$S^0(t_1) \cdots S^0(t_n)$
in the object initial state 
is approximately equal to the $n$th moment of $S$,  
for any $0 \leq t_1,\cdots ,t_n  \lesssim \dectime, t_{\rm int}$.
Similarly, the free dynamics of $\Pc$ is ineffective  to move the pointer
between $t=0$ and $t \approx t_{\rm dec}, t_{\rm int}$. 
Note that $\TS=\infty$ if $S$ commutes with $\HS$. The conditions on $\TS$ in
(\ref{timescales}) are necessary in order that an eigenstate $\ket{s}$
of $S$ be left almost unchanged by the
measurement. The conditions on $\TP$ 
(in particular, the
high-temperature limit $\hbar \beta \ll \TP$) would be difficult
to avoid for a macroscopic pointer.

A further key input
is  the quantum central limit theorem
(QCLT)~\cite{QCTL} which implies Gaussian statistics
(Wick theorem)
for the bath coupling agent in the pointer-bath interaction, as
discussed below.
The limits (\ref{timescales}) and the QCLT imply validity of our
results for a wide class of objects and apparatus. An extreme
case of universality~\cite{Haake01} arises under the further
stipulation that entanglement and decoherence be faster than the decay
of bath correlations (time constant $\TB$). Indeed, for 
$\dectime,t_{\rm int}\ll \TB$ even the free bath motion remains ineffective during
the measurement.  Opposite to that slow-bath limit is the Markov
regime $\TB \ll \dectime, t_{\rm int}$. Our analysis below covers
both these regimes as well as the intermediate case 
$\dectime , t_{\rm  int} \approx \TB$.


\section{The model}  
\subsection{Object, pointer and bath}

We consider a
three-partite system comprising the microscopic object $\Sc$, a
single-degree-of-freedom macroscopic pointer $\Pc$ and a bath $\Bc$ with many
($N\gg 1$) degrees of freedom (labelled by $\nu$).
The following dynamical variables 
 will come into play: for 
$\Sc$, the
observable ${S}$ to be measured, assumed to have a discrete
spectrum; 
for $\Pc$, the position ${X}$ and
momentum ${P}$; and for $\Bc$, a certain coupling agent $B$.  The
pointer is coupled to $\Sc$ and $\Bc$ by the Hamiltonians
\begin{equation} \label{eq-int_Hamiltonian}
\HPS = \epsilon  {S} {P}
\;\; , \;\;
\HPB = {B} {X}
\;\; ,\;\;{B}=N^{-1/2} \sum_{\nu=1}^N B_\nu\,. 
\end{equation}
The object-pointer coupling Hamiltonian $\HPS$ is chosen so as to
(i)~not change the measured observable $S$ (i.e., $[\HPS,S]=0$);
(ii)~be capable of shifting the pointer position by an amount
proportional to ${S}$, such that each eigenvalue $s$ of $S$ becomes
tied up with a specific pointer reading; (iii)~involve a large
coupling constant $\epsilon$, so that different eigenvalues $s\neq s'$
end up associated with pointer readings separated by large distances.
The pointer-bath interaction is chosen for most efficient decoherence
of distinct pointer positions~\cite{Haake01}; the 
additivity of $B$  in
contributions $B_\nu$ acting on single degrees of freedom of the
bath will
allow us to invoke the QCLT.

Let us point out an essential difference between our model 
and the interacting
spin model of~\cite{Balian03}. Unlike in this reference,
$\Sc$ is strongly coupled to a single degree of freedom 
(the pointer $\Pc$) of the apparatus, e.g. with its total momentum
$P$ in a given direction. The coupling of
$\Sc$ with the other apparatus degrees of freedom (the bath $\Bc$, for us)      
is assumed to be much weaker and can therefore be neglected (see~\cite{long_paper}).  

The full Hamiltonian is $\HPSB = \HS+\HP+\HB + \HPS + \HPB$.  We need
not specify $\HS$. The pointer Hamiltonian $\HP=P^2/(2M)+V(X)$ must
allow for a well-defined rest state.  
We assume that $V(x)$ has  a local minimum at $x=0$, i.e., 
$V'(0)=0$ and $V''(0)>0$.  
The time scale for free pointer motion then is the  period
$\TP=2\pi(M/V''(0))^{1/2}$ of oscillations around this minimum.  
Like the coupling agent $B$, the bath
Hamiltonian $\HB$ is assumed additive as
$\HB=\sum_{\nu} H_{\Bc,\nu}$ with
 $[ H_{\Bc,\nu},B_\mu]\propto \delta_{\mu \nu}  \dot{B}_\mu$ 
and $[H_{\Bc,\mu},H_{\Bc,\nu}]=0$, 
again clearing ground for the QCLT.

%
\subsection{Initial state: pointer localized around $x=0$, apparatus
 in thermal equilibrium} 

It is appropriate to require initial statistical
independence of the object and the apparatus, and thermal equilibrium
for the apparatus. The initial density operator $\rhoS(0)$ of the
object may represent a pure or a mixed state.  The full initial
density operator reads
\begin{equation} \label{eq-initial_state}
\rhoPSB (0) = \rhoS (0) \otimes  \rhoPB(0) \, , \qquad 
\rhoPB(0) = Z_{\cal{PB}}^{-1} \,
\E^{-\beta (\HP + \HB + \HPB)}\;.\label{eqapp}
\end{equation}
By invoking the high-temperature limit $\hbar \beta \ll \TP$ and the Gaussian
statistics of $B$ (as implied by the QCLT) and by tracing out the bath
we find (see below) the reduced  density matrix
$\rhoP(0)=\trB\rhoPB(0)$ of the pointer in  the position representation,
\begin{equation} \label{eq-rhoP(0)}
\bra{x}\rhoP(0) \ket{x'} \propto
\E^{-\beta (V_{\rm eff}(x)+V_{\rm eff}(x'))/2}\,
\E^{-2 \pi^2 (x-x')^2/\lamb_{\rm th}^2}
\end{equation}
where $\lamb_{\rm th}=2\pi\hbar(\beta/M)^{1/2}$ is the thermal de
Broglie wavelength.  The pointer potential here appears renormalized
by the pointer-bath interaction as
\begin{equation} \label{eq-effective_pot}
V_{\rm eff}(x)
= V(x)-(\gamma_0/\hbar)x^2 
\, , \qquad 
\gamma_0=\int_{-\infty}^{0} \D t  \,
\Im h(t )\,\geq 0
\end{equation}
where $h(t)$ is the autocorrelator of the bath coupling agent  
with respect to
the free bath thermal  state $\rhoB \propto \E^{-\beta\HB}$,
\begin{equation} \label{eq-h(t)}
h(t) 
= \langle \tilde{B}(t)B\rangle_0 = 
\tr_{\Bc} \tilde{B}(t)B \rhoB \, ,\qquad 
\tilde{B}(t)=\E^{\I\HB t/\hbar}B\E^{-\I\HB t/\hbar}
\end{equation}
and we assume $\langle B \rangle_0 = 0$.  For stability of the
whole apparatus the pointer-bath coupling must be
weak enough so that $V''_{\rm eff}(0)>0$; we even bound the latter
curvature finitely away from zero by, say,  
$V_{\rm eff}''(0)>V''(0)/2$,  i.e.,
\begin{equation} \label{eq-stability}
\gamma_0 /\hbar < V''(0)/4\,.
\end{equation}
 This makes sure that the
initial density $\bra{x} \rhoP (0) \ket{x}$ of pointer positions has a
single peak at $x=0$ with a renormalized width $\Delta_{\rm eff} = [
\beta (V''(0) - 2 \gamma_0/\hbar) ]^{-1/2}$ of the order of the bare
thermal fluctuation $\Delta_{\rm th}=(\beta V''(0))^{-1/2}$.

If $V(x)={\rm o} (x^2)$ at large distances $|x|$,
the effective potential $V_{\rm eff}(x)$ is unstable.
The matrix elements (\ref{eq-rhoP(0)}) then correspond to (the reduced
pointer state of) a {\it local thermal equilibrium}.
That local equilibrium for the apparatus can be achieved by 
preparing $\Pc$ in some state localized near $x=0$ at time
$t=-t_{\rm i}$ and then letting it
interact with $\Bc$ between $t=-t_{\rm i}$ and $t=0$. 
If  the thermalization time is small compared with
the tunnelling escape time, one may choose
$t_{\rm i}$ larger than the former but
much smaller than the latter time, so that
$\Pc$ is still within the
effective potential well when the measurement starts at $t=0$.
In order to be able to prepare the apparatus in 
such a local equilibrium,
the height  $V_0^{\rm eff}$ of the two potential barriers 
surrounding the local minimum of 
$V_{\rm eff}(x)$ at $x=0$ must be 
large compared with the thermal energy $\beta^{-1}$. 
Thanks to (\ref{eq-stability}), this is the case provided that 
the bare potential $V(x)$ satisfies the same requirement, 
i.e., $V_0 \gg \beta^{-1}$.
Interestingly,  $V(x)$ can be chosen 
such  that the two potential barriers of $V_{\rm eff}(x)$ 
are separated by a mesoscopic distance 
$W_{\rm eff} \approx (V_0^{\rm eff}/ V_{\rm eff}''(0))^{1/2} \gg 
\Delta_{\rm eff}$ (so that $V_0^{\rm eff} \gg \beta^{-1}$) 
which is small compared with  the
macroscopic read-out scale $\Delta_{\rm class}$.
 The object-pointer
interaction then just has to get the pointer out of the well,
leaving the subsequent displacement growth to the action of
the effective pointer potential. The  instability resulting from the 
pointer-bath coupling (\ref{eq-int_Hamiltonian}) 
hence provides an amplification mechanism. For a
macroscopic pointer at high temperature ($\hbar \beta \ll \TP$), the
different length scales  are ordered as
$\lamb_{\rm th} \ll \Delta_{\rm th} \approx \Delta_{\rm eff} \ll
W_{\rm eff} \ll \Delta_{\rm class}$.
%


\section{Object-pointer dynamics}  


Assuming that the  bath state is not
ascertainable, we  define
the object-pointer density matrix at time $t$ as
$\rhoPS(t)=\trB \E^{-\I H t/\hbar} \rhoPSB(0) \,\E^{\I H t/\hbar}$.
Accepting a relative error $\Or(t/\TP,\,t/\TS)$, 
we simplify the time evolution operator at time $t \ll \TS,\TP$ as
$\E^{-\I Ht/\hbar} \simeq U(t)\,\E^{-\I(\HS+\HP)t/\hbar}$, with
\begin{equation}
\fl
\qquad U (t)  
= \E^{-\I  (\HB+\HPS +\HPB)t/\hbar}
= \E^{-\I\HB t/\hbar} \E^{-\I\epsilon S Pt/\hbar}
\left(\E^{-\I\int_0^t \D \tau (X+\epsilon S\tau)\tilde{B}(\tau)/\hbar}\right)_+
\, .
\end{equation}
Here, $(\cdot )_+$ denotes time ordering; the momentum $P$ was used as
the generator of pointer displacements, $\E^{\I\epsilon S P t
  /\hbar}X\E^{-\I\epsilon S P t /\hbar}=X+\epsilon S t$.  
Similarly, we invoke $\bra{s,x} \E^{-\I\epsilon S P
  t/\hbar}=\bra{s,x-\epsilon s t}$, the cyclicity of the trace 
and the product initial state (\ref{eq-initial_state}) 
to get the matrix elements of
$\rhoPS(t)$ 
in the joint eigenbasis $\{ \ket{s,x} \}$ of $S$ and $X$, 
\begin{equation}
\nonumber
\bra{s,x}\rhoPS(t) \ket{s',x'} \simeq  
\bra{s} \rhoS^{0} (t)\ket{s'}\;\bra{x}\rhoP^{ss'}(t)\ket{x'}
\quad , \quad t \ll \TS,\TP\, ,
\end{equation}
with 
\begin{equation} \label{eq-rhoS^0}
\rhoS^{0} (t)=\E^{-\I \HS t/\hbar}
\,\rhoS (0) \, \E^{\I \HS t/\hbar}
\end{equation}
evolving as for the free
object, while the pointer matrix elements
\begin{equation} \label{eq-pointer_matrix_elements}
\bra{x}\rhoP^{ss'}(t)\ket{x'}
=
  \bra{x_s(t)}
  \trB\tilde{U}_{sx}(t)\rhoPB(0)\tilde{U}_{s'x'}(t)^\dagger
  \ket{x_{s'}'(t)}
\end{equation}
 involve the bath
evolution operator and shifted positions 
\begin{equation} \label{eq-x_s(t)}
\fl
\quad 
 \tilde{U}_{sx}(t)  =  \left(\E^{-\I\int_0^t d\tau\, x_s(t-\tau)
      \tilde{B}(\tau)/\hbar}\right)_+ \,, \quad 
x_s(t) = x- \epsilon s t\,,
\quad x_{s'}'(t) = x'- \epsilon s' t \, .  
\end{equation}
Note that $\E^{-\I\HP
  t/\hbar}\rhoPB(0)\,\E^{\I\HP t/\hbar} \simeq \rhoPB(0)$ for $t \ll
\TP$. 
The Hamiltonian $\HS$ cannot be neglected in (\ref{eq-rhoS^0}), even for $t
\ll \TS$, although  for such times 
$\bra{s} \rhoS^{0}(t) \ket{s} \simeq \bra{s}\rhoS \ket{s}$.

To evaluate the matrix elements
(\ref{eq-pointer_matrix_elements}) we use the high-temperature
approximation 
\begin{equation}
\nonumber
\rhoPB(0)\simeq Z_{\cal{PB}}^{-1} \,
\E^{-\beta \HP/2} \,\E^{-\beta (\HB + \HPB)}\E^{-\beta \HP/2} 
\end{equation}
for the pointer-bath Gibbs state (\ref{eqapp}). 
Given the weak-coupling condition 
$ \Delta_{\rm  th}^2 \beta^2\langle B^2\rangle_0 < 1/2$ 
implying the stability (\ref{eq-stability}) (thanks to
$\gamma_0 \leq \hbar \beta \langle B^2\rangle_0/2$, see~\cite{long_paper}), 
the error incurred is
$\Or(\hbar^2\beta^2/\TP^2)$, as easily 
seen from the Baker-Campbell-Haussdorf formula. Thus 
\begin{equation} \label{eq-y-integral}
\fl
  \bra{x}\rhoP^{ss'}(t)\ket{x'}
\propto 
\int  dy\,
  \bra{x_s(t)}\E^{-\beta \HP/2}\ket{y}
  \bra{y}\E^{-\beta \HP/2}\ket{x'_{s'}(t)}
\, Z_{{\cal B},y}\,
  \big\langle\tilde{U}_{s'x'}(t)^\dagger\tilde{U}_{sx}(t)
  \big\rangle_{y}
\end{equation}
where $\langle\cdot\rangle_{y}$ is the bath average with respect to
the modified equilibrium state $\rho_{{\cal B},y}=Z_{{\cal B},y}^{-1}
\,\E^{-\beta(\HB+yB)}$; the normalization factor  
 is determined by using the QCLT as~\cite{long_paper}
\begin{equation} \label{eq-Z_By}
Z_{{\cal B},y}=\E^{\beta\gamma_0 y^2/\hbar}
Z_{{\cal B},0}\,.
\end{equation}

At this point we momentarily pause with dynamics and show that at
$t=0$, when $\tilde{U}_{s'x'}=\tilde{U}_{sx}=1$, $x_s=x$ and
$x_{s'}'=x'$, (\ref{eq-y-integral}) yields the initial pointer
state announced in (\ref{eq-rhoP(0)}). To that end we invoke high
temperatures $\hbar \beta \ \ll \TP$ again to approximate the matrix
element $\bra{x}\E^{-\beta\HP/2}\ket{y}$ by $\E^{-\beta (V(x)+V(y))/4}
\E^{-4\pi^2(x-y)^2/\lamb_{\rm th}^2}$. 
Replacing $V(y)$ by $V''(0) y^2/2$ in that expression 
and doing the Gaussian $y$-integral in (\ref{eq-y-integral}), 
we arrive at the initial state
(\ref{eq-rhoP(0)}) by neglecting terms $\Or( \lambda_{\rm
  th}^2/\Delta_{\rm th}^{2}, \lambda_{\rm th}^{2}/\Delta_{\rm
  eff}^{2})$.

Let us return to the time-evolved pointer matrix (\ref{eq-y-integral}).
Since $\rho_{{\cal B},y}$ factors into single-degree-of-freedom states, the QCLT
assigns Gaussian statistics to the bath coupling agent $B$ for the average
$\langle \cdot\rangle_{y}$, with a mean
$\langle\tilde{B}(\tau)\rangle_{y}\propto y$ given by linear response
theory and a variance {\it independent of
  y}~\cite{long_paper},
\begin{equation} \label{eq-varianceB}
\fl\quad
\langle \tilde{B}(t)\rangle_{y}=-\frac{2y}{\hbar}
\int_{-\infty}^t d\tau \,\Im h(\tau)
\, , \quad
\langle \tilde{B}(t) \tilde{B}(t') \rangle_{y}
-\langle \tilde{B}(t)
\rangle_{y} \langle \tilde{B}(t') \rangle_{y} = h(t-t') \,.
\end{equation}
Therefore,
$\langle\tilde{U}_{s'x'}(t)^\dagger\tilde{U}_{sx}(t)\rangle_{y}$
coincides with its value for $y=0$ up to a phase factor
$\E^{\I\int_0^t\!d\tau \langle\tilde{B}(t-\tau)\rangle_{y}
  (x'_{s'}(\tau) -x_s(\tau))/\hbar}$ 
and the $y$-integral in (\ref{eq-y-integral}) remains Gaussian for
$t>0$.

Of prime importance is the {\it decoherence factor} 
\begin{equation}
\nonumber
\langle\tilde{U}_{s'x'}(t)^\dagger\tilde{U}_{sx}(t)\rangle_0 = 
\E^{-D_t(x_s(t),x_{s'}'(t);s,s')-\I\phi_t(x,x';s,s')}
\end{equation}
with a {\it positive decoherence exponent}  $D_t$ 
 revealed by the QCLT as~\cite{Haake01,long_paper}
\begin{eqnarray} \label{eq-D_t}
\nonumber 
\fl\quad
D_t(x,x';s,s')=\frac{1}{\hbar^2} \int_0^t d\tau_1
\int_0^{\tau_1} d\tau_2 \,\Re h(\tau_1-\tau_2)  
\\ 
\qquad \times
\big(x'_{s'}(-\tau_1)-x_s(-\tau_1)\big)
                  \big(x'_{s'}(-\tau_2)-x_s(-\tau_2)\big)
\end{eqnarray}
and a  real phase $\phi_t$ irrelevant for decoherence. 
Let us stress that the aforementioned results (in particular (\ref{eq-D_t})) 
are exact 
(not lowest order in the pointer-bath coupling). They are 
consequences of Wick's theorem as
  implied by the  QCLT and the additivity (\ref{eq-int_Hamiltonian}) 
of the bath  coupling agent $B$. Direct proofs of (\ref{eq-Z_By}),
(\ref{eq-varianceB}) and  (\ref{eq-D_t})  are easy
in the particular case of a bath composed of
harmonic oscillators linearly coupled to $\Pc$.

It turns out~\cite{long_paper} that 
the phase factor
$\langle\tilde{U}_{s'x'}(t)^\dagger \tilde{U}_{sx}(t)\rangle_{y}
/\langle\tilde{U}_{s'x'}(t)^\dagger \tilde{U}_{sx}(t)\rangle_{0}$\,%
entails nothing but a correction of relative order 
$(\lamb_{\rm  th}/\Delta_{\rm eff})^2$ 
to the decoherence exponent $D_t$ under the stability condition
(\ref{eq-stability}).
Dropping that correction, the $y$-integral reduces to the initial
pointer density matrix (\ref{eq-rhoP(0)}), albeit with the shifted
pointer positions $x\to x_s(t) = x-\epsilon st$ and $x'\to
x'_{s'}(t) = x'-\epsilon s't$ reflecting the action of the
object-pointer coupling. Our final result for the object-pointer state
at time $t \ll \TS,\TP$~is
\begin{equation} \label{eq-object-pointer-state}
\fl  \qquad 
\bra{s,x}\rhoPS(t)\ket{s',x'}
= \bra{s} \rhoS^0(t)\ket{s'}\;
\bra{x_s(t)}\rhoP(0)\ket{x'_{s'}(t)}
\E^{-D_t(x_s(t),x'_{s'}(t);s,s')-\I\phi_t}
\end{equation}
with the notations specified
in  (\ref{eq-rhoP(0)}), (\ref{eq-rhoS^0}) 
and (\ref{eq-D_t}).
Entanglement and 
decoherence contribute separately in  that
remarkably simple ``final state'';  
they lead respectively to  
the second and third factors in
(\ref{eq-object-pointer-state}). 
The decoherence (third) factor equals unity for $s=s'$ and $x=x'$.

\section{Discussion}  

\begin{figure}[ht]
\centering
\psfrag{D}{$D^{\rm peak}_t$}
\psfrag{ln(T)}{$\ln \tau_{\rm dec}$}
\psfrag{ln(t)}{$\ln (\tau_{\rm ent}/\eta)$}  
\psfrag{t}{$\tau$}
\psfrag{tdec}{$\tau_{\rm dec}$}
\includegraphics[width=7cm]{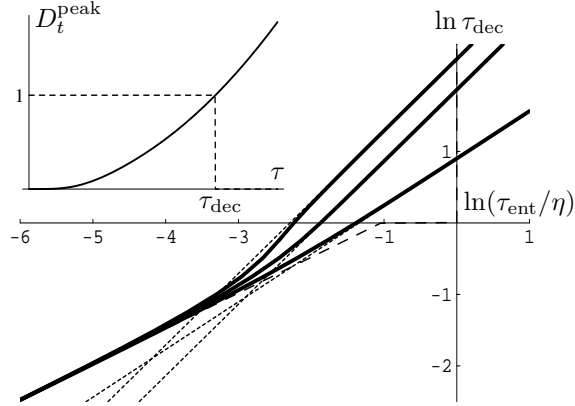}
\caption{\label{fig-lett}
Decoherence time 
$\tau_{\rm dec}=t_{\rm dec}/\TB$ 
in units of $\TB$  against $\enttime/(\eta\,\TB)$ 
in a $\log$-$\log$ scale. We take
$\widehat{\Re h}(\omega)= \I \coth (\hbar \beta\omega/2) \widehat{\Im
  h}(\omega)$ (KMS relation) and 
$\widehat{\Im h}(\omega) \propto \omega^m \E^{-(\hbar \beta
  \omega/5)^2}$. 
The larger decay time of 
the bath correlator $h(t)$ is then the thermal time $\TB=\hbar \beta$.  
Solid curves: $m=5,3,1$ (from left to right). 
Broken curves: 
approximate expressions
(\ref{t_dec_short_time}-\ref{D_t-Markov-Ohmic})
for $\tau_{\rm dec} \ll 1$ (dashed lines) and  $\tau_{\rm dec} \gg 1$
(dotted lines). 
Inset: decoherence exponent $D^{\rm peak}_t$
against $\tau=t/\TB$ ($m=3$).
}
\end{figure}

Given the narrow peaks (of
width $\Delta_{\rm eff}$) of the initial pointer density matrix
(\ref{eq-rhoP(0)}) at $x=x'=0$, one can appreciate the fate of the $s\neq
s'$ coherences in the  final state (\ref{eq-object-pointer-state}) by
setting $x_s(t)=x'_{s'}(t)=0$ there. The decoherence factor then reads 
\begin{equation} \label{decexp}
\E^{-D_t^{\rm peak}(s,s')}
= 
\exp 
\left\{  
 -\frac{\epsilon^2(s-s')^2}{\hbar^2} \!\!
   \int_0^t \! \D \tau_1 \! \! \int_0^{\tau_1} \! \D \tau_2 \, \tau_1 \tau_2 \,
    \Re h(\tau_1-\tau_2)
\right\}
\end{equation}
and reveals irreversible decay as soon as the time $t$ much exceeds
the {\it decoherence time} $t_{\rm dec}(s,s')$; we may define that time
implicitly as $D_{t_{\rm dec}}^{\rm peak}(s,s')=1$.
It can be shown~\cite{long_paper} that $D_t^{\rm peak}(s,s')$ is
an increasing convex function of time 
(see the inset in figure~\ref{fig-lett}).

The diagonal ($s=s'$ and $x=x'$) terms in the final state
(\ref{eq-object-pointer-state}) give the probability density of pointer positions for
fixed $s$. That density has a sharp peak (of width $\Delta_{\rm eff}$)
at
$x=\epsilon st$. The peaks associated to distinct
$s$ and $s'$ begin to be resolved  at the {\it entanglement time}
$t_{\rm ent}(s,s')= \Delta_{\rm eff}(\epsilon |s-s'|)^{-1}$. That time 
is related to $\dectime(s,s')$ by
\begin{equation}
\label{eq-dectime_vs_enttime}
\left( \frac{\enttime (s,s')}{\eta} \right)^2 
=
\frac{1}{(\hbar\beta)^2} \int_0^{\dectime (s,s')}\D \tau_1
\int_0^{\tau_1} \D \tau_2 \, \tau_1 \tau_2\, \frac{\Re h (\tau_1
  -\tau_2)}{\langle B^2\rangle_0}
\end{equation}
 where $\eta=\langle B^2\rangle_0^{1/2}\Delta_{\rm eff}\beta$ is a
dimensionless measure of the strength of the pointer-bath coupling.
Figure~\ref{fig-lett} shows $\dectime$ as function of 
$\enttime/\eta$
for three distinct choices of the bath correlator $\Re
h(t)/\langle B^2\rangle_0$.


Let us note that, in analogy with the results of~\cite{Balian03},
the $s\not= s'$ matrix elements
of the reduced  density matrix of $\Sc$, $\bra{s} \tr_{\Pc} \rhoPS (t)
\ket{s'}$, 
decay to zero on a time scale
$\lambda_{\rm th} (\epsilon |s-s'|)^{-1}$ much shorter than both
$\enttime(s,s')$ and $\dectime(s,s')$ (see (\ref{eq-rhoP(0)}) 
and (\ref{eq-object-pointer-state})).

Recalling that $S$ has a discrete spectrum,
we
 denote by $\ds$ the minimum of $|s-s'|$ over all 
pairs $(s,s')$ of eigenvalues present in the object 
initial state (we suppose that 
$\bra{s} \rhoS \ket{s'} =0$ if $s$ and $s'$ belong to
a part of the spectrum containing arbitrarily close
eigenvalues, near an accumulation point, so that $\ds >0$). 
At time $t > \enttime= \Delta_{\rm  eff}(\epsilon \ds)^{-1}$, 
neighbouring peaks 
of the pointer densities  can be resolved. 
Each eigenvalue $s$ of the measured object observable is then
uniquely tied up with ``its'' pointer position $\epsilon st$.  
If $t$ is also much larger than the maximum decoherence time    
$\dectime=\dectime(s,s\pm \ds)$
($t$ being still smaller than $\TS$ and $\TP$),
the  matrix elements
(\ref{eq-object-pointer-state}) for $s\not= s'$
almost vanish
for all values of $(x,x')$. 
Assuming moreover that the spectrum of $S$ is non-degenerate, 
object and pointer are 
in a separable
mixed state, $\rhoPS(t)\simeq \sum_s p_s \ketbra{s}{s} \otimes
\rhoP^{s s}(t)$, with $p_s = \langle s | \rhoS^{0}(t)| s \rangle
\simeq \langle s | \rhoS | s \rangle$. 
Hence, according to the Born rule, with probability $p_s$ the object
is in the eigenstate $\ket{s}$ and the pointer is in a state 
$\rhoP^{s s}(t)$ localized near
$x=\epsilon s t$ with probability density $\bra{x} \rhoP^{s
  s}(t)\ket{x} \propto e^{-\beta V_{\rm eff} (x - \epsilon s t)}$, 
in agreement with the von Neumann postulate.
The coupling $\HPS$ may be switched off at time $t_{\rm
  int} \approx W_{\rm eff} (\epsilon \ds)^{-1} \gg \enttime$, where
$W_{\rm eff}$ is defined in Subsection 2.2. Then all pointer
states $\rhoP^{ss}(t)$ 
 are outside the effective potential well 
save for one eigenvalue $s=0$. 
The inter-peak distance 
is amplified at time $t > t_{\rm int}$ by the
effective
pointer dynamics, till it reaches 
 a macroscopically resolvable magnitude $\Delta_{\rm class}$.
Then a pointer reading, while still a physical process in principle
perturbing $\Pc$, surely cannot blur the distinction of the peaks.

\section{Limiting regimes} 


Formula (\ref{eq-dectime_vs_enttime}) explicitly
yields the decoherence time in two interesting limits. An {\it
  interaction dominated} regime has decoherence outrunning bath
correlation decay ($\dectime\ll\TB$) such that we can use
$h(\tau)\simeq \langle B^2\rangle_0$ in (\ref{decexp}) and 
(\ref{eq-dectime_vs_enttime}). We conclude
\begin{equation} \label{t_dec_short_time} 
\fl \qquad
\E^{-D_t^{\rm peak}
    (s,s')}
= \E^{-(t/\dectime(s,s'))^4}
\, ,   \qquad
\frac{\dectime (s,s')}{\hbar\beta}
= 2^{3/4}\left(
    \frac{\enttime (s,s')}{\hbar \beta\,\eta } \right)^{1/2} 
\end{equation}
for $\dectime\ll\TB,\TS,\TP$.  
The decoherence time depends on the bath through
the pointer-bath coupling strength $\eta$ only.
It is smaller than $t_{\rm int}$ when 
$\enttime \geq 8^{1/2} \eta^{-1} ( \Delta_{\rm eff}/W_{\rm  eff})^2 \hbar \beta$. 
The small-time behaviour (\ref{t_dec_short_time}) of the 
decoherence factor appears also in other models~\cite{Balian03}.

The opposite limit $\dectime\gg\TB$ defines the {\it Markovian regime}. 
A rotating-wave approximation is inappropriate due to our restriction (\ref{timescales}).
Assuming that the Fourier transform of  the imaginary part of $h$
behaves as $(\widehat{\Im h})(\omega)\propto \omega^m$ when $\omega \rightarrow 0$,
we find
\begin{equation} \label{D_t-Markov-Ohmic}
\fl \qquad 
\E^{-D_t^{\rm peak} (s,s')} 
= \E^{-(t/\dectime(s,s'))^\gamma} 
\, , \qquad 
 \frac{\dectime (s,s')}{\hbar\beta}
= c_m^{1/\gamma}
    \left( \frac{\enttime (s,s')}{\hbar \beta\,\eta } \right)^{2/\gamma}
\end{equation}
for  $\TB \ll \dectime\ll\TS,\TP$. Here $\gamma=3$
for a ``Ohmic bath'' ($m=1$) and
$\gamma=2$ for a ``super-Ohmic bath'' ($m\geq 3$); the constant
$c_m$ is independent of the strengths of the couplings (\ref{eq-int_Hamiltonian}),
$c_{1}=3 \hbar \beta \langle B^2\rangle_0/ \int_0^\infty \D \tau \Re h
(\tau)$ and
$c_{m\geq 3} = 2 \hbar^2
\beta^2 \langle B^2\rangle_0/|\int_0^\infty \D \tau\,\tau\, \Re h (\tau)
|$.

In
all cases the coherences decay non-exponentially.  
We compare in figure~\ref{fig-lett} 
the asymptotic results
(\ref{t_dec_short_time})-(\ref{D_t-Markov-Ohmic}) with numerical
solutions of (\ref{eq-dectime_vs_enttime}). 
One finds an excellent agreement except for a thin range around $\dectime =
\TB=\hbar \beta$.
If $\eta \lesssim 1$,  
the only regime with a decoherence faster
than resolution of pointer peaks 
($\dectime \leq \enttime$) is the Markov regime ($\enttime \geq
 c_1 \eta^{-2} \hbar \beta$)  
with $m=1$ (Ohmic bath).  
In all asymptotic regimes $\dectime \leq t_{\rm int}$
 if $\eta \geq c_m^{1/2} \Delta_{\rm eff}/W_{\rm eff}$ and 
$\enttime \geq \hbar \beta \max\{ (8/c_m)^{1/2} ,1 \} \Delta_{\rm eff}/W_{\rm
  eff}$. 
This means that mesoscopic superpositions decay to mixtures
faster than entanglement can create them.

\section{Conclusion and outlook}

We have investigated
a model for quantum detection in which the entanglement produced
by the coupling of the measured object with the pointer
is simultaneous with decoherence of distinct pointer readouts; the
apparatus (pointer and bath) is taken initially in a local thermal  metastable
state, not correlated to the object.
We have shown that the decoherence time $\dectime$
presents a universal behaviour in the interaction-dominated
regime $\dectime \ll \TB$, whereas it depends strongly
on the small-frequency behaviour of
the bath correlator  in the Markov regime $\dectime \gg \TB$. 
For  reasonably
strong pointer-bath and weak 
object-pointer
couplings, 
$\dectime$ is smaller than  
the time $t_{\rm int}$ needed by entanglement to produce 
mesoscopic superpositions, which do not appear at any time
during the
measurement.

Several  generalizations of our results present no
difficulties. First, nonlinear pointer-bath couplings, $\HPB=B X^\alpha$ 
with $\alpha>1$, make for richer decoherence scenarios and 
produce smaller decoherence times save for Ohmic baths in the Markovian
regime~\cite{long_paper}.  Second, the QCLT also works for
baths of interacting particles if the correlator $\langle B_\mu B_\nu
\rangle_0$ decays more rapidly than $1/|\mu-\nu|$ (see~\cite{Verbeure}
for a related version of the QCLT in this context). 
We shall publish these elaborations elsewhere.

\newpage

\ack

We acknowledge support by the Deutsche Forschungsgemeinschaft (through
the project 
SFB/TR 12) and the Agence Nationale de la Recherche (project
ANR-05-JCJC-0107-01) and thank M.  Guta for pointing out~\cite{QCTL} to us.

\section*{References}





\end{document}